# Interface Features and Users' Well-Being: Measuring the Sensitivity of Users' Well-Being to Resource Constraints and Feature Types

Oded Nov

Tandon School of Engineering, New York University, New York, NY, USA, onov@nyu.edu

**ABSTRACT**
Users increasingly face multiple interface features on one hand, and constraints on available resources (e.g., time, attention) on the other. Understanding the sensitivity of users' well-being to feature type and resource constraints, is critical for informed design. Building on microeconomic theory, and focusing on social information features, users' interface choices were conceptualized as an exchange of resources (e.g., time), in return for access to goods (social information features). We studied how sensitive users' well-being is to features' type, and to their cost level and type. We found that (1) increased cost of feature use leads to decreased well-being, (2) users' well-being is a function of features' *cost type*, and (3) users' well-being is sensitive to differences in *feature type*. The approach used here to quantify user well-being derived from interface features offers a basis for asynchronous feature comparison.

**Author Keywords**
Feature economics; social information; user preferences; user interface

**CSS Concepts**
• **Human-centered computing~Human computer interaction (HCI)**; *HCI theory, concepts and models*

## 1. Introduction

In their interactions with user interfaces, users increasingly face multiple interface features (e.g., sorting, filtering, ratings) on one hand, and constraints on their available resources (e.g., time, attention, screen real estate) on the other. The relationship between the two and the tradeoffs they represent for these users, however, are far from understood. A fundamental question which can serve as a starting point for unpacking this relationship – and which received surprisingly little attention in the literature – concerns the nature of users' well-being derived for user interface features: to what extent a certain UI feature adds to its users' well-being, given the constraints placed on the user's resource? Understanding and quantifying users' feature preferences and sensitivity to resource constraints is critical for informed design: many user tasks take place in environments in which users' resources are constrained [3, 4, 24], and as a result, designers have to make various design decisions, such as which types of information they want to present to users, and at what level of detail. These decisions can be better informed if they are based on an understanding of users' preferences and constraint sensitivities.

This paper reports the use of a novel approach to studying users' well-being derived from user interface features. It applies microeconomic theory and tools to understand and quantify user choices, and it treats UI features as economic goods: users' choices are conceptualized as informed transactions, in which users engage in an exchange of scarce resources (e.g. time), in return for access to goods (UI features – e.g., sort or filter). Thinking about features as economic goods (products or services that can be obtained in return for constrained resources) is useful as a mechanism with which to think about tradeoffs in the allocation of resources of different UI stakeholders [35] – that is, both the designer's perspective (by addressing questions such as which features are more cost-effective in invoking desired user behavior), and the user's perspective (by addressing questions such as which features are more desirable, and what constraints their use impose on users). From a design practice perspective, thinking about interface features as economic goods and focusing on the tradeoffs between them as denominated in scarce resources, offers a basis for asynchronous comparison between different features in different circumstances. reducing the need to A/B test every small set of design alternatives, a feature economics approach acts in much the same way as monetary prices of economic goods – it frees designers from the need to directly compare consumer preference for every set of alternative products, the same product in different consumer circumstances, or both.

In the present study, we investigate the well-being users derive from using interface features. Economists has long measured changes in consumers' well-being by observing their *consumer surplus*, the difference between the their willingness to pay for a good and the amount that they actually pay [5]. With this approach, and echoing a growing interest in and calls for more use of economic theory in the study of HCI [1, 17, 28, 35, 39, 45] and digital goods [5], this paper makes a number of contributions, by advancing the feature economics research agenda: first, UI features are conceptualized as economic goods to which users attribute a quantifiable value, operationalized as the scarce resources, or costs (e.g., time, money), they are willing to incur. Users' choices are conceptualized as informed transactions, in which users engage in an exchange of scarce resources in return for access to goods (UI features). Second, a method for eliciting users' consumer surplus for features is established, which can be applied across different settings, use cases and resource constraints. As a result, findings from three controlled experiments are presented in which the method is used to elicit consumer surplus as a function of three factors: cost level, cost type, feature type. Finally, the findings contribute to the understanding of the role of HCI in personal finance [17, 19, 20, 25, 48, 50], and to the research agenda bringing economic theory to the study of HCI [3, 9, 17, 27, 35, 44, 49].

## 2. Related work

Applying economic theory to the study of users' choices in the context of interface design alternatives, Toomim et al [45] used the concepts of utility and revealed preference [30] – the idea that agents' preferences can be revealed and compared by observing their choices between multiple goods. Viewing utility as the extent to which a user prefers a particular choice over others, and considering all factors of functions and usability that affect users' preference and use, Toomim et al [45] suggested that *Utility(A) > Utility(B)* when a user chooses to use interface A instead of interface B. Building on this notion, they developed a framework for eliciting a crowdsourcing labor supply curve given different interface variations. Using an online crowdsourcing tasks, they investigated the relationships between task price, UI aesthetics and ease of use on worker supply. In a similar vein, other researchers [7, 13, 22, 42, 43] discussed, developed and tested methods and tools to determine crowdsourcing pricing, and in particular, efficient and fair pricing of crowdsourcing tasks. Other studies used economic theory concepts to study the relationship between design, behavior and utility, by focusing on the availability of different features to users, for example, by examining users' visual search strategies used given online images' ecology constraints [46]. Prior research, however, tended to focus on the supply side of online work in the context of different user interfaces, with little attention to the users' demand side – the choices users make about interface features' use as they go about performing their tasks.

Building on the notion that user interface preferences can be elicited experimentally, prior research examined the feasibility of eliciting demand for a *sort* feature [35], a commonly used UI feature. While this prior work constructed the feature's demand curve relative to its cost, the findings were limited to one type of cost (monetary) and one type of feature (sort), and are therefore silent on how *variation* in cost and feature type affect demand.

Various types of social information features were studied in prior research: some involve direct interaction or other financial and informational exchange between users, some include the presentation of "feeds" of information about other users, and others include the ability to broadcast information to large audiences of unknown others. Specifically, HCI research studied many settings in which the social aspect of the study was the effect of information about the choices or behaviors of other users on users' own choices or perceptions, with examples including the effects of information about others on investment decisions [50], trust and expertise perceptions [23], impression of peer workers [29], participation in distributed crowd work [26], aggregated recommendations in recommender's systems and users' agreement with their output [6, 10, 14, 32, 34, 51] and others.

In the context of user behavior and personal finance, prior work examined how people manage and think about their money [25, 37, 48], how they make decisions about saving [17], and how financial information can be packaged in novel ways to assist with comparing potential investments [18]. In recent years, personal finance is becoming increasingly social, with the emergence of peer-to-peer lending and payments platforms such (e.g. Prosper) which combine enable monetary exchange informed by social information. Understanding the value people place on social information is essential to the understanding (and design) of social systems that bring people to exchange not only information or information good (as they do in Facebook or Twitter), but also money (which social media platforms also attempt to do). Prior studies, however did not consider a comparison of users' well-being associated with feature use, the tradeoff that features' choice represents, and the patterns of user surplus derived from social information features.

## 3. Research Questions

A first step in eliciting users' surplus for interface features builds on prior work [11, 35, 46, 47] and focuses on users' well-being derived from the use of a given interface feature. To that end, the tradeoff between the extent to which a user is interested in access to a UI feature, and the feature's cost, needs to be elicited quantitatively. To operationalize this tradeoff for a user, we need to observe the user's consumer surplus - the constraints the user is willing to incur in order to gain access to the feature. In other words, we need to check if features are similar to most economic goods, in that a tradeoff exists between how much the user wants to use them and the scarce resources they are willing to forego in order to do that.

RQ1: How sensitive users' surplus is to a feature's *cost*?

A second question concerns the sensitivity of the well-being derived from a feature to the feature's cost *type* – how what the constraint on the user's resources is. Consider the example of time and money, both of which are scarce resources which are often treated as costs that people budget for: people may be highly sensitive to differences in the cost of an airline ticket, if the cost to them is denominated in money, but much less sensitive to the differences if the cost they face is denominated in the duration of the layover. In other words, they may prefer a flight that is 1% cheaper than an alternative flight, but would be less sensitive to a 1% difference in layover time (assuming both price in money and layover time are cost people want to minimize). We can therefore say that in this example, the sensitivity to cost (the tradeoff between cost and likelihood of choosing an alternative) is a function of the cost *type*.

RQ2: How sensitive users' surplus is to a feature's *cost type*?

A third research question concerns differences in the well-being derived from a feature, resulting from the differences in the type of the feature: since there are, potentially, many possible feature types designers can incorporate into a user interface to serve a given functionality, some variation is likely to exist in the degree to which users' choices are influenced by such design choice.

RQ3: How sensitive users' surplus is to differences in *feature type*?

In the present study we focus on social information features that facilitate the presentation of information about other users' aggregated choices – which in other settings may include the number of Likes a social media post receives, the number of responses to stories on news outlets, or products' popularity rankings on e-commerce sites. We are interested in how such social information help users make decisions when they are presented with a large number of available alternatives to choose from.

## 4. Method

To ensure high ecological validity and relevance to design, a useful context in which to study users' interaction with different features and under different constraint types involves choice overload [8] – a situation in which people face many alternatives to choose from, which can in turn lead to multiple undesired consequences including users' paralysis, users' poor choice-making, and users' dissatisfaction with choices, even good ones [36]. Prior research focused on multiple strategies to reduce choice overload, including the use of social information as a common heuristic for dealing with choice overload [40], variation of the choice set size [36], and the ranking of results [8]. Echoing a long tradition in economic research, we used discrete choice experiments to observe users' consumer surplus [5].

To address the research questions, a series of experiments was conducted using the setting of personal finance. HCI research is increasingly involved in the study of user behavior in personal finance, examining how people think about and interact with their money [19, 25, 35, 37, 44, 48, 50]. The personal finance context offers a number of advantages: first, interaction with financial information often requires users to limit the number of alternatives they consider, and use information provided online as they make investment choices. When making investment decisions, users face hundreds (and in some cases thousands) of alternatives they can potentially select from, and which require assessing tradeoffs between attributes of these alternatives, such as the potential risk, reward and other factors. This is a natural setting in which to study choice overload and the information people turn to in order to reduce the overload. Second, since extant research shows that choice overload when considering financial alternative products impacts consumers' choices [2, 31, 41], often leading to reduced consumer welfare (e.g. not saving enough for retirement), understanding choice overload can have an important societal impact.

## 5. Experimental procedure

To elicit a range of user surplus data reflecting different constraint and feature types, randomized experiments were developed in which users were tasked with making multiple performance-incentivized investment and feature use choices: after completing a short demographic survey, users were asked to go through ten investment rounds. In each round they had a fixed budget they could invest in any combination of investment alternatives, using up to 77 different funds (see Figures 1 and 4). The large number of alternative possible investment choices presented to users was intentionally designed to induce choice overload [2, 31, 41], which is typical in many online environments, ranging from e-commerce to movie rating to dating platforms [12]. Since considering all 77 alternatives and their tradeoffs is highly time consuming, a natural heuristic for dealing with such choice overload would be for a user to seek additional information about the alternatives they face. Information about the choices of other users has been shown to have a strong effect on users' own choices, as they implicitly assume that the popularity of an alternative reflects its quality [21, 40].

The funds users could invest in were provided fictional names, but were based on commercial financial products such as stock and bond funds, representing different levels of risk and reward. This is a realistic scenario, similar to many investment environments people face when they save. Funds' performance in the experiment was based on real historic market data, but this was unknown to the users. Every round, users allocated their budget across funds, and once they submitted their allocations, were informed of the return on their investment. Users then proceeded to the next round.

As users made their investment allocation decisions, interface features were available to them upon activation, presenting the average allocations to the 77 funds made by previous participants, in that same round. This information was based on real users' data obtained from the users' choices in a control condition (no intervention) in a prior study [35].

| Funds | Category | Annual fee (%) | Price | Average annual return: 1 year | Average annual return: 5 years | Average investment percentage allocated by other participants | Allocation |
|---|---|---|---|---|---|---|---|
| CM0062 | Money Market | 0.16% | $ 1.0 | 3.29% | 0.56% | 0.83% | 0% |
| MT0168 | Bond - Long-term State Muni | 0.15% | $ 10.8 | 4.91% | 0.96% | 0.94% | 0% |
| LI0723 | Balanced | 0.12% | $ 15.4 | 7.93% | 0.99% | 0.54% | 0% |
| MG0026 | Stock - Large-Cap Growth | 0.38% | $ 28.4 | 11.09% | 1.07% | 0.38% | 0% |
| DM0127 | International | 0.07% | $ 13.5 | 26.27% | 1.53% | 2.92% | 0% |
| E00024 | Stock - Small-Cap Growth | 0.46% | $ 95.7 | 9.7% | 1.18% | 0.6% | 0% |
| SF0049 | Bond - Short-term Government | 0.20% | $ 10.7 | 4.32% | 0.79% | 0.81% | 0% |
| CS0082 | Balanced | 0.34% | $ 13.3 | 12.94% | 1.22% | 0.62% | 0% |
| LT0024 | Bond - Short-term | 0.10% | $ 11.0 | 3.00% | 0.74% | 0.04% | 0% |

**Figure 1. Investment allocation page: Users were presented with 77 alternative funds to which they could allocate their budget. Users who choose to activate it, can view the column in which the average allocation made by previous participants to each fund is presented.**

To elicit consumer's surplus for an economic good, economic theory suggests to observe the difference between consumers' willingness to pay for a good and the amount that they actually pay [5]. In cases such as user interface features, many of which are free, this would involve setting a cost denominated in a constrained resource, the user will have to forgo in order to achieve the good. This way, the person needs to choose between the good and some other resources, and the cost they are willing to incur reflects the utility of that good to them. While prior studies focused on the positive and negative value of aesthetics [1] and users' annoyance [16], in the present study the choice of operationalizing cost and benefit reflected the task users were engaged in: maximizing their investment return in a choice overload situation, given their time constraints. Therefore, identifying the time constraints people are willing to take on, or the amount of money they are willing to forgo, represents a practical and ecologically valid way to study how users value features in a financial decision making context.

Participants in all three experiments were recruited through Amazon Mechanical Turk. Participation was limited to U.S. users with a record of at least 100 tasks at an approval rate above 95%. To incentivize users to make realistic choices, their compensation was primarily based on investment performance, consisting of a basic participation fee of $1.5, and on top of it, a bonus that was calculated as a percentage of users' total investment return throughout the ten rounds. If the return on the investment was negative, no bonus was given in addition to the base pay. In the experiments, the bonuses ranged between 0-$3.99, and the average net bonus was $1.2. This compensation structure represents a substantial incentive for participants to pay attention to their investment alternatives and consider ways to enhance their return.

**EXPERIMENTS**
To address the research questions, we conducted three experiments, following the procedure described above. Reflecting the study design, the experiments incorporated variations of feature and constraint type and level (see Table 1):

|  |  | Variation in constraint type | |
|---|---|---|---|
|  |  | Monetary constraint | Time constraint |
| Variation in feature type | Sorting with detailed information | Experiment 1 | Experiment 2 |
|  | Sorting with detailed vs. summarized information |  | Experiment 3 |

Table 1. The experimental design integrating variations in feature types and resource constraints.

**Experiments 1-2: Method and Participants**
Experiments 1 and 2 addressed the first two research questions, concerning the tradeoff between feature cost and use at different cost levels, and across different cost types. Users were randomized into multiple cost levels and two cost types: money (Experiment 1) and time (Experiment 2). In both cases the effects of variation in feature cost on feature use were examined.

The general experimental design enabled users to activate the feature by incurring a cost of either money or time: in Experiment, 1, money was used to elicit the user surplus associated with feature use. Users were presented a dialog box at the beginning of each round, informing them that they could to forego a fraction of their bonus from the round they were in (that is, the bonus to be given to them if they make a positive return on their investment) in order to activate the feature, for that round. The size of the bonus fraction users could forego in order to access the feature ranged between 0%-9% of their bonus for that round. The percentage presented to users were randomly assigned every round, such that users had to consider different costs in different rounds. The relationship between the likelihood that a user will activate the feature and the cost in money they had to incur to do so was then analyzed. 148 users, whose average age was 36.2 and of whom 43.3% identified as women, participated in this experiment (each of them going through 10 rounds).

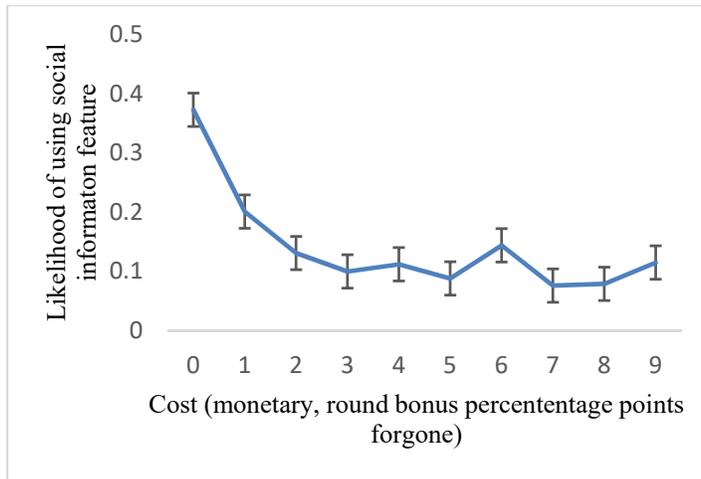

**Figure 2. Demand for the feature as a function of its cost, in bonus fraction forgone. Error bars represent standard error of the mean.**

A variation of this approach, reflecting variation in constraint type, and focusing on the utility of time, was used in Experiment 2. Here, the constraint users could forego in return for access to the interface feature, was denominated in users' time. Each round users were offered to put a limit on the amount of time available to them in order to activate the feature. The length of the time limit offered to users varied, taking randomly one of the values of 20, 40, 60, 80, 100 and 200 seconds. The time limit offered to users in return for activating the feature was randomized each round. Users were presented a dialog box at the beginning of each round, in which they could either take on the time limit for making their decision and gain access to the feature, or take as long as they want to make their decision, but with no access to the feature. The option of no time limit was set in order to make Experiment 2 as close as possible to Experiment 1, and with the objective to contrast and price a constraint (in time or money) with a no-constraint alternative. To make sure users dedicated at least a minimal amount of time for weighing their choices, the analysis included investment rounds longer than ten seconds. Here too, the relationship between the likelihood that a user will activate the feature and the cost (in time) they had to incur to do so, was analyzed. 117 users participated in this experiment. Their average age was 34.7 and 38.9% identified as women.

**Experiments 1-2: Results**
In Experiment 1, a logistic regression was used to examine the relationship between the likelihood that users will activate the feature, and the related costs, in monetary terms. The analysis resulted in significant differences in feature use across cost levels (Wald=37.78, $p<0.0001$), a downward sloping curve (see Figure 2) and curvilinear relationship between the feature cost and the likelihood of its use. That is, the likelihood that a user will choose use the feature was lower the more costly the feature was to them. However, this sensitivity to increases in cost flattened beyond a cost level of 2%. Next, it was important to rule out the possibility that the differences in feature activation stem from users' reluctance to pay for the feature. Using a logistic regression, we compared the likelihood that users will activate the feature across different cost points, limiting the analysis the costs higher than zero. We found a significant difference (Wald=6.02, $p=0.014$) between the levels of this likelihood, indicating that users are sensitive to cost beyond just to the notion of paying a positive amount for feature activation.

In Experiment 2, a logistic regression was used to examine the relationship between the likelihood that users will activate the feature, and the related costs, this time denominated in the time constraints users chose to take on in return. The analysis of users' choices and cost data revealed an upward sloping curve (see Figure 3) and significant differences between the likelihoods of the feature use as a function of its cost level (Wald=10.72, $p=0.001$). That is, in this experiment too, users' likelihood to use the feature was higher the less costly the feature was to them. Note that the x axis, representing the cost, starts on the left from high cost (more constraint on available time), unlike the x axis in the case of monetary cost which starts on the left from low cost (in Figure 2). In this experiment, since the cost was denominated in time constraint, a cost of zero was not possible and a time constraint of 200 seconds a round was used as a substitute for a very low cost.

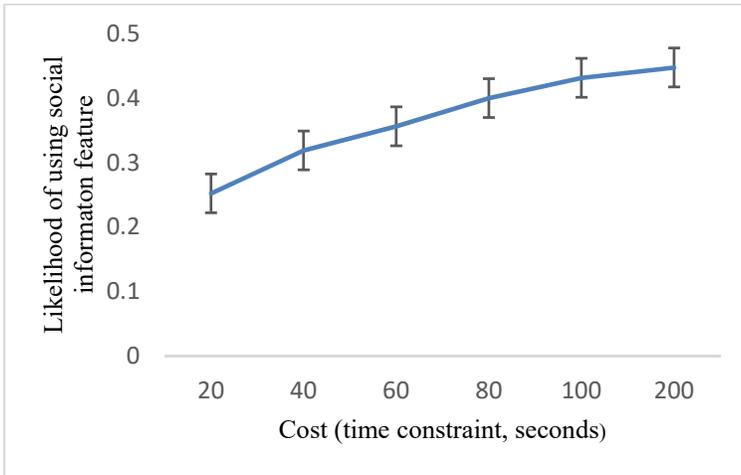

Figure 3. Demand for the feature as a function of its cost in self-imposed time constraint (where shorter time represents higher cost). Error bars represent standard error of the mean.

**Experiments 1-2: Discussion**
A consistent pattern was found of decreasing surplus significantly associated with an increase in the cost of access to the feature, for both cost types, which is consistent with the pattern for most economic goods [15]. However, the findings of experiments 1 and 2 also suggest that users' sensitivity to the cost of is a function of the cost type: in the monetary cost case (Experiment 1), users' surplus became virtually insensitive to cost beyond a relatively low cost point (2% and more – see Figure 2). This result demonstrates a similar pattern of flattening surplus, but a greater sensitivity to cost compared to direct (not social) information gathering UI features found in previous studies, where the demand for sorting was found to flatten when its cost exceeded 4% of the bonus [35]. In comparison, when users' cost was denominated in time constraints (Experiment 2), the surplus decreased across all cost levels (see Figure 3). These findings complement prior studies on the value of information (e.g., [38]) in which users reveal their preferences at different price points.

| | Money | Stocks | Balanced | International | Bonds |
|---|---|---|---|---|---|
| | 4.81% | 46.5% | 7.3% | 14.3% | 26.9% |

| Funds | Category | Annual fee (%) | Price | Average annual return: 1 year | Average annual return: 5 years | Allocation |
|---|---|---|---|---|---|---|
| PM0053 | Stock - Sector Precious Metals and Mining | 0.43% | $ 10.2 | 34.3% | 2.22% | 0% |
| PA0559 | Stock - Large-Cap Growth | 0.33% | $ 124.0 | 12.49% | 1.2% | 0% |
| HT0044 | Bond - Long-term National Muni Index | 0.19% | $ 11.3 | 5.53% | 1% | 0% |
| SI0548 | Stock - Small-Cap Blend Index | 0.06% | $ 64.9 | 15.77% | 1.38% | 0% |
| VI0506 | Stock - Large-Cap Value | 0.06% | $ 37.5 | 22.27% | 1.27% | 0% |
| GI0509 | Stock - Large-Cap Growth | 0.06% | $ 66.7 | 9.13% | 0.81% | 0% |

Figure 4. Investment allocation page with an alternative social information feature: users who choose to activate it, see a summary of the top five asset categories chosen by other users in each round.

**Experiment 3: Method and Participants**
In this experiment, RQ3 was addressed. To study the effect of feature type on users' choices, a new feature was designed and implemented, presenting to the user a summary of how other participants allocated their investment across the five asset types in each round, rather than across specific funds (see Figure 4). Since many funds presented to users are fairly similar, a summary overview of the asset types invested in by others, offers a broad overview of the detailed information presented in the feature of Experiments 1 and 2.

The key difference between the information presented in Experiment 3 and the information presented in Experiments 1 and 2, is that other users' average choices were now grouped into asset categories and presented as a summary of others' choices. From a design perspective, this is a summary vs. detailed presentation of the information. With a summary view, the choice overload users faced with the detailed view of others' choice was expected to be reduced. While there may be many other features that the original feature could have been compared to, the goal here was to examine a plausible alternative, that offers a different perspective – rather than different data – on the same information presented in Experiments 1 and 2. The cost of activating the feature was denominated in the time constraint users could take on. 63 users participated in this experiment. Their average age was 35.1 and 32.5% identified as women.

**Experiment 3: Results**
The analysis of users' choices whether or not to use the information summary feature as a function of its cost, revealed a similar pattern to that of the detailed information feature: an upward sloping curve (see Figure 5, lower curve). A logistic regression revealed significant differences (Wald=20.66, p<0.0001) in the likelihoods of the feature use across cost levels.

The results also show a difference in surplus, with the likelihoods of choosing to use the feature significantly different across both feature type and cost level (Wald=16.18, p<0.0001 for the former and 20.47 p<0.0001 for the latter). Reflecting the regression results, Figure 5 presents the results of both experiments 2 and 3, and enables a comparison of the two feature types in which cost was denominated in time constraints.

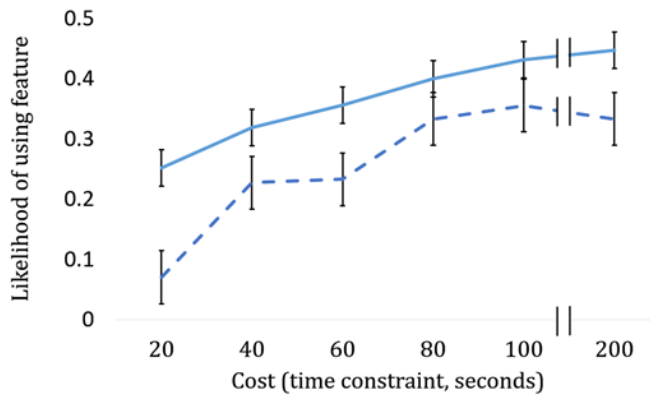

**Figure 5. Comparative surplus for two features: others' allocation - detailed (solid line), and others' allocation – summary (dashed line). Error bars represent standard error of the means.**

**Experiment 3: Discussion**
In this experiment a comparison between the surplus associated with different feature types was sought, by examining the tradeoff between feature activation cost and use across different feature types. The comparison provides insights into how users value different aggregation types of information (which was the key difference in this experiment). In our case, the detailed view of the information proved to be perceived as more valuable by others (that is, users were willing to "pay" a higher price for it).

Beyond the results of this particular experiment, the findings, using a constrained resource as the currency users for feature activation, offer a demonstration of a method for comparison of features with future features, that have yet to be developed or implemented. Establishing methods and offering empirical insights about the comparative surplus patterns associated with interface features is important for researchers and practitioners who operate in an environment in which the multiple features potentially available for designers.

## 6. General Discussion and Conclusion

Can we characterize quantitatively users' well-being associated with features, and the sensitivity to the constraints placed on users' resource? A method to address such questions can be highly valuable to researchers and practitioners: many user tasks take place in resources constrained contexts, and therefore designers have to make decisions about the types of features and information to be included in UIs, and the level of detail they represent. Such decisions can be better informed when based on clear understanding of users' preferences and constraint sensitivities.

To demonstrate the feasibility of such characterization, in the experiments presented here we elicited user surplus, focusing on the use of social information interface features in situations of choice overload [40]. It should be noted, however, that users' surplus associated with an interface feature is highly dependent on the task the user is engaged in. For example, a user who is interested in a quick answer to a time-sensitive question, would probably value speedy access to information in an aesthetically displeasing website, compared to a user who is interested in learning about a topic with no specific question in mind and less pressing time constraints. Understanding and quantifying such tradeoffs can help in creating an environment that caters to users' needs.

The focus on the role of interface features on users' behavior in a choice overload situation required a use case in which such choice overload is natural. Personal finance offers a useful setting, since, in their interaction their finances, users often face more information than they can comprehend effectively, and need to reduce the number of alternatives they can consider. By conceptualizing interface features as economic goods, and by varying the constraint types and levels users may "pay" with, we could experimentally elicit users' surplus function. This function, and in particular, a comparison of the surplus across interface features, is highly useful for designers weighing design choices tradeoffs [35].

To elicit the surplus for features realistically, participants were incentivized by making their compensation performance-based. Prior work was extended by using two different cost types – money (in Experiment 1) and time (in Experiment 2). The surplus elicitation experimental design was then used to demonstrate (in Experiment 3) how to compare users' surplus to two different information types. A similar procedure, in which feature and cost types vary in an otherwise-similar environment and then compared using a regression analysis in which both are independent variables, can be applied when comparing any number and types of information features.

From a design practice perspective, the work presented here offers a way to study the return on investment in UI design [9]. Moreover, by examining two cost types and feature types, the approach used in the experiments makes such investments more evidence-based, informed, and explainable. Being able to quantify user surplus as a function of feature costs denominated in time constraints can help designers make informed decisions as they consider the amount of time users are expected to spend on a given web page or app, and given the users' goals in visiting it. Being able to quantitatively compare surplus across feature types is another way to inform designers' choices: using the surplus elicitation procedure applied in the experiments enables designers to compare the extent to which users value features in comparison to other present or future features. As such, it enables designers to make asynchronous comparisons, in a sequential and iterative process, and reduces the need for simultaneous A/B tests of a finite set of alternatives as means of feature evaluation. While there are many other features and cost types one can compare, the approach presented here offers a guideline on how to develop such comparisons. Beyond the specifics of the results presented here, generalizing the findings to other types of settings and features would enable us to gain deeper insights into users' preferences and the tradeoffs they make, explicitly or implicitly, as they use interface features.

Future work could address the limitations of the work presented in this paper. First, the choice of using the personal finance context should be considered with the understanding of its limitations: being asked to perform a task within a personal finance scenario may prime users to think about the value of their interactions with user interface features primarily in monetary terms. Moreover, in a crowd work context, users are likely to consider alternative earnings potentially available by working on other tasks, and therefore prioritize their time constraints. These characteristics of the task used in this work may therefore implicitly direct users' attention to costs denominated in money or time. For a more generalizable study of surplus patterns across cost types, alternative, constraint types can be examined in future research. Costs can be operationalized in different ways, including variations of factors that have been studied in prior research – e.g. aesthetics [1] and user annoyance [16]. Second, while only two feature types, in one user setting, were studied in this paper, future work can explore other features and other user settings and use cases. Third, the experimental procedure presented here did not include a training phase, which could have helped users calibrate their time value estimates. It should be noted, however, that the effect of the training phase absence is likely to exist for all users, and therefore the patters of surplus found throughout the experiments would persist even with the presence of a training session. To check this assumption, we ran additional analyses of the experiments' data, this time eliminating the first round's results (that is, running the analyses only for rounds 2-10). We found that the use patterns (as a function of feature cost) were not different from the original experiments, therefore alleviating concerns about the lack of an experiment training phase. Fourth, the studies presented here did not account for differences between users in terms of their individual income constraints and risk aversion – both of which may impact users' choices.

With the growing constraints placed on users' available resources, including time, attention and screen real estate on one hand, and calls for evidence-based design on the other, the HCI community can benefit from understanding, developing and using feature economics. More informed use of methods that have been developed to study allocation of scarce resources and people's choices [33], can help researchers and practitioners make design decisions and quantify the implications of these decisions.